\documentclass[3p,times,procedia,twocolumn,twoside]{elsarticle}
\usepackage{ecrc}
\usepackage{amsmath}
\usepackage{bm}
\usepackage[figuresright]{rotating}

\volume{00}

\firstpage{1}

\journalname{Procedia CIRP}

\runauth{Santiago Arroyave-Tob\'on et al.}

\jid{cirp}

\biboptions{sort&compress}

\usepackage{amssymb}

\begin{document}

\begin{frontmatter}

\dochead{14th CIRP Conference on Computer Aided Tolerancing (CAT)}

\title{Adapting polytopes dimension for managing degrees of freedom in tolerancing analysis}

\author[*]{Santiago Arroyave-Tob\'on}
\author[]{Denis Teissandier}
\author[]{Vincent Delos}

\address{Univ. Bordeaux, I2M, UMR 5295, F-33400 Talence, France}

\begin{abstract}
In tolerancing analysis, geometrical or contact specifications can be represented by polytopes. Due to the degrees of invariance of surfaces and that of freedom of joints, these operand polytopes are originally unbounded in most of the cases (i.e. polyhedra). Homri et al. proposed the introduction of virtual boundaries (called cap half-spaces) over the unbounded displacements of each polyhedron to turn them into 6-polytopes. This decision was motivated by the complexity that operating on polyhedra in $\mathbb{R}^6$ supposes. However, that strategy has to face the multiplication of the number of cap half-spaces during the computation of Minkowski sums. In general, the time for computing cap facets is greater than for computing facets representing real limits of bounded displacements. In order to deal with that, this paper proposes the use of the theory of screws to determine the set of displacements that defines the positioning of one surface in relation to another. This set of displacements defines the subspace of $\mathbb{R}^6$ in which the polytopes of the respective surfaces have to be projected and operated to avoid calculating facets and vertices along the directions of unbounded displacements. With this new strategy it is possible to decrease the complexity of the Minkowski sums by reducing the dimension of the operands and consequently reducing the computation time. An example illustrates the method and shows the time reduction during the computations.
\end{abstract}
 
\CorText[cor1]{Corresponding author. Tel.: +33-5-4000-8790; fax: +33-5-4000-6964. \email{santiago.arroyave-tobon@u-bordeaux.fr}}
\belowfrontmatterskip20pt
\end{frontmatter}


\setlength{\belowdisplayskip}{15pt} 
\setlength{\belowdisplayshortskip}{15pt}
\setlength{\abovedisplayskip}{15pt} 
\setlength{\abovedisplayshortskip}{15pt}

\enlargethispage{-13mm}

\section{Introduction}

The objective of tolerance analysis is to determine if the cumulative defects fulfil the functional requirements of a mechanical system. The displacements limits of a toleranced surface inside its tolerance zone or a toleranced joint inside its clearance can be modelled by a set $P$ of $n$ half-spaces $\bar{H}^-_i =  \left \{ a_{i1}x_1 + ... + a_{i6}x_6  \leq b_i \right \}$ in $\mathbb{R}^6$ \cite{teissandier_operations_1999,teissandier_algorithm_2011}:
\begin{align*}
   P = \bigcap_{i=1}^n \bar{H}^-_i &= \{\bm{x} \in \mathbb{R}^6: \bm{a_i}^T x \leq b_i, i=1,...n \} \\
 &=\{\bm{x} \in \mathbb{R}^6: \bm{Ax}\leq \bm{b}\}, \bm{A}\in \mathbb{R}^{n\times 6}\     \text{and}\ \bm{b} \in \mathbb{R}^n
\end{align*}

The six dimensions are due to the six possible displacements that define the position and orientation of any rigid body with respect to a coordinate system in the Euclidean space. Since the degrees of invariance $d_{inv}$ of toleranced surfaces or the degrees of freedom $d_{mob}$ of toleranced joints define theoretically unbounded displacements, $P$ is usually an open set, i.e. a polyhedra in $\mathbb{R}^6$.

Due to the complexity that operating on polyhedra in $\mathbb{R}^6$ supposes, Homri et al. \cite{Homri2015103, Homri2015112} proposed the introduction of virtual boundaries $\bar{Hc^-}_j$, called cap half-spaces, over the unbounded displacements of geometric or contact polyhedra to turn them into 6-polytopes. Finally, $P$ becomes a bounded set $\mathcal{P}$ by adding $m=2.d_{inv}$ half-spaces for a toleranced surface or $m=2.d_{mob}$ for a toleranced joint.
\[
\mathcal{P} = \left( \bigcap_{i=1}^n \bar{H}^-_i \right) \cap \left( \bigcap_{j=1}^m \bar{Hc}^-_j \right)\ 
\]

Once each set of constraints becomes a bounded set, the accumulation of variations in a mechanical assembly can be calculated through Minkowski sums and intersections in $\mathbb{R}^6$ \cite{teissandier_algorithm_2011,delos2015Minkowski}.

\enlargethispage{-13mm}
This strategy, suitable even for tolerance analysis of over-constrained assemblies, has to face the multiplication of cap half-spaces during the computation of Minkowski sums. Thus, the time for computing cap facets (facets associated with cap half-spaces) is far greater than that for computing facets representing real limits of bounded displacements.

In order to deal with the aforementioned issue, we propose the use of the theory of screws to perform the mobility analysis of the toleranced mechanical system. By doing so, the set of bounded displacements that define the relative position of a couple of surfaces influencing the functional condition (FC) can be determined. For each case, this set of displacements defines the smallest affine subspace of $\mathbb{R}^6$ in which the respective polytopes have to be projected to perform the Minkowski sum. As a result, the complexity of this highly time-consuming operation can be decreased by reducing the dimension of the space.

\section{Tolerance analysis with 6-polytopes}
\label{sec:six_polytopes}

\begin{figure}[htbp]
	\centering
		\includegraphics[width=5.5cm]{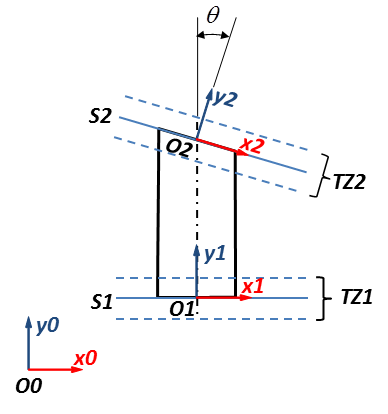}
	\caption{Proposed case study}
	\label{fig:Case_study}
\end{figure}

The current methodology of tolerance analysis with 6-polytopes is illustrated in this section by an example. The case presented in figure \ref{fig:Case_study}, implies the control of the relative position of two non-parallel nominally planar surfaces $S_1$ with a local reference system $R_1$ (vectorial base $x_1$, $y_1$ and $z_1$) and $S_2$ with a local reference system $R_2$ (vectorial base $x_2$, $y_2$ and $z_2$) and tolerance zones $TZ_1$ and $TZ_2$ respectively. \\

\begin{figure}[ht]
	\centering
		\includegraphics[width=6.5cm]{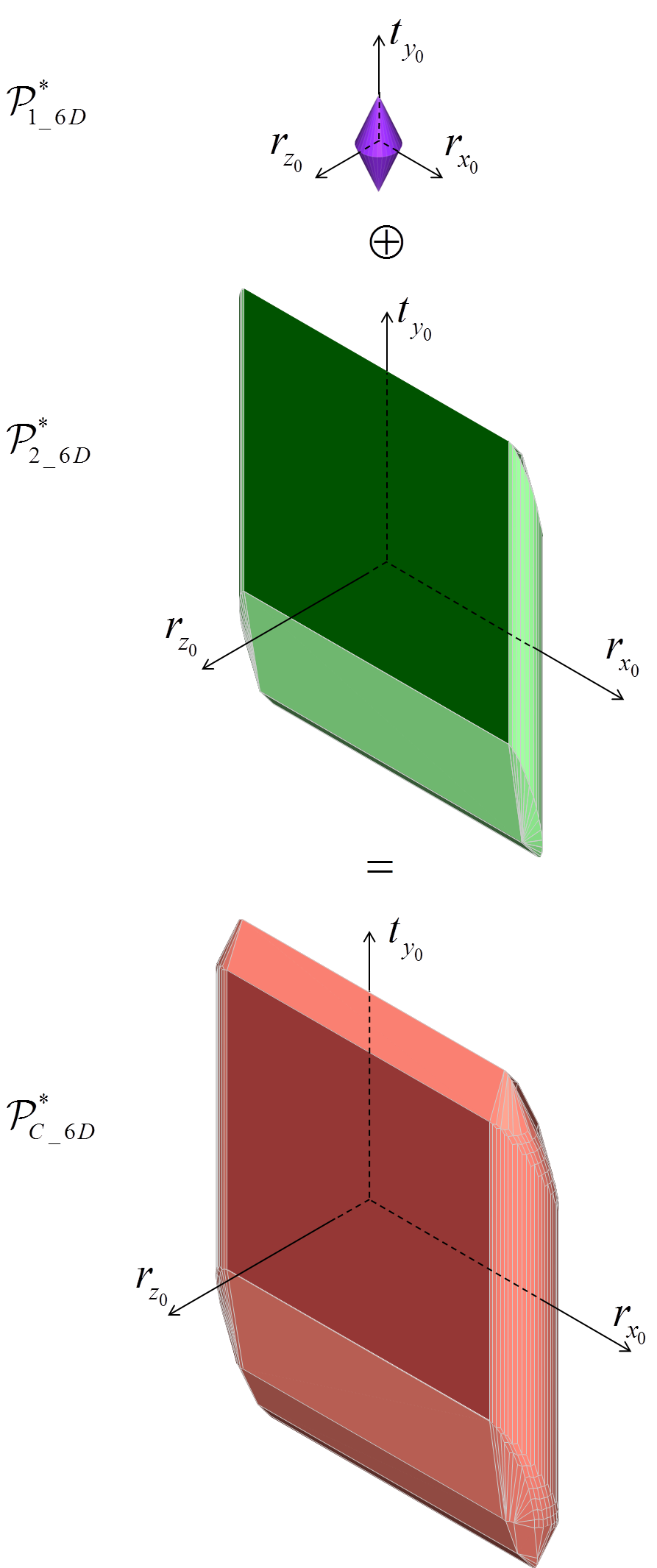}
	\caption{3D projection of $\mathcal{P}_{1\_6D}$, $\mathcal{P}_{2\_6D}$ and $\mathcal{P}_{C\_6D}$. Facets highlighted in darker color represent non-cap facets}
	\label{fig:sum6d}
\end{figure}

To solve the proposed case with the 6-polytopes method, first a CAD file representing the toleranced part was created and imported into the open source software PolitoCAT \cite{PolitoRef}. The software, through its graphical interface, allows the creation of 6-polytopes representing possible variations of toleranced features. The 6-polytope $\mathcal{P}_{1\_6D}$, representing the possible displacements of $S_1$ inside its tolerance zone $TZ_1$, was created with 20 pairs of geometric constraints, generated in turns by a discretization of the contour line in 20 points. To virtually limit the unbounded displacements of $S_1$, 3 pairs of cap half-spaces were added to the set of geometric constraints. Particularly for this case, the cap half-spaces are required to limit the rotation along $y_1$, $r_{y_1}$, the translation about $x_1$, $t_{x_1}$, and the translation about $z_1$, $t_{z_1}$. $O_1$ was chosen to express the constraints. As it is not possible to represent graphically 6-polytopes, a projection into a 3D space is required. The projection of $\mathcal{P}_{1\_6D}$, $\mathcal{P}_{1\_6D}^{*}$, is shown in figure \ref{fig:sum6d}, where the axis of projection are the rotation along $x_0$, $r_{x_0}$, the rotation along $z_0$, $r_{z_0}$ and the translation along $y_0$, $t_{y_0}$.
Similarly, the 6-polytope $\mathcal{P}_{2\_6D}$ representing the possible displacements of $S_2$ inside its tolerance zone $TZ_2$ was created with 20 pairs of geometric constraints and 3 pairs of cap half-spaces. The 3D projection $\mathcal{P}_{2\_6D}^{*}$ of $\mathcal{P}_{2\_6D}$ onto $r_{x_0}$, $r_{z_0}$ and $t_{y_0}$ is presented in figure \ref{fig:sum6d}. It can be noted that $\mathcal{P}_{2\_6D}^*$ has unbounded displacements along $r_{x_0}$ and $t_{y_0}$ due to the influence of the unbounded displacements $r_{y_2}$ and $t_{x_2}$ in the local base of the surface. This is why much of the facets of $\mathcal{P}_{2\_6D}^{*}$ become cap half-spaces in the global base. Figure \ref{fig:sum6d} shows in darker color the non-cap facets.

The strategy based on 6-polytopes propose the direct sum of $\mathcal{P}_{1\_6D}$ and $\mathcal{P}_{2\_6D}$ in $\mathbb{R}^6$. This operation was performed by means of the open source software politopix \cite{PolitoRef}. The operation took 19 s to be computed, the details of the calculated polytope $\mathcal{P}_{C\_6D}$ and its operands are presented in table \ref{tab:6dresults}. The simulation was performed with an Intel Core i7-3740QM. Figure \ref{fig:sum6d} shows a 3D projection $\mathcal{P}_{C\_6D}^{*}$ (according to $r_{x_0}$, $r_{z_0}$ and $t_{z_0}$) of the sum $\mathcal{P}_{C\_6D} = \mathcal{P}_{1\_6D} \oplus \mathcal{P}_{2\_6D}$.

By analyzing the graphical results it can be concluded that the only controllable displacement is $r_{z_0}$. It is worth to mention that for facility of visualization of the polytopes, the second member of the cap half-spaces was chosen not too big, but the graphical difference between the bounded and unbounded displacements is even clearer when this value is increased. This conclusion is also compliant with the trace of the bounded displacements of the operands: for the operand $\mathcal{P}_{2\_6D}$ the only bounded displacement in the global base ($x_0$,$y_0$,$z_0$) is $r_{z_0}$. Then, the unbounded displacements of this operand are kept in the calculated polytope $\mathcal{P}_{C\_6D}$.

\begin{table}[h!]
\centering
\caption{Tolerance analysis with 6-polytopes (F: facets, V: vertices)}
\label{tab:6dresults}
\begin{tabular}{llll}
\hline
             & $\mathcal{P}_{1\_6D}$    & $\mathcal{P}_{2\_6D}$ & $\mathcal{P}_{C\_6D}$ \\ \hline
F            & 46       & 46    & 3676   \\
V            & 176      & 176   & 5208   \\
Time {[}s{]} &          &       & 19   \\ \hline
\end{tabular}
\end{table}

These results show that computing Minkowski sum of polytopes in $\mathbb{R}^6$, implies the computation of many facets coming from cap half-spaces of the operands, which were initially required just to generate closed sets. Punctually in this example, the calculated polytope in $\mathbb{R}^6$, $\mathcal{P}_{C\_6D}$, is composed by 3676 facets, from which just 2 are necessary to describe the tolerance analysis problem: the two bounding the displacements along the axis $r_{z_0}$. All the remaining facets are coming from cap half-spaces and have no meaning in this tolerance analysis problem. In other words, polytopes in $\mathbb{R}^1$ are enough for solving this case. Hence the question: is it possible to know the bounded displacements (the dimension) of the calculated polytope before performing a Minkowski sum? The answer can be found in the mobility analysis of the toleranced surfaces. This can be carried out by means of the theory of screws as it is explained in the next section.


\section{Theory of screws overview}
\label{sec:theory_of_screws}
The theory of screws is based on the following theorems \cite{ball1900treatise}:
\begin{itemize}
\item Chasles' theorem: any rigid body motion can be represented instantaneously as a rotation about a unique line and a translation along that same line.
\item Poinsot's theorem: any system of moments and forces acting on a rigid body can be represented instantaneously as a one moment and one force.
\end{itemize}

These theorems describe the concepts of twist and wrench respectively. Twists can be analyzed as allowable motions while wrenches as forbidden motions \cite{Su2013Type}. Both, twists $\bm{\hat{T}}$ and wrenches $\bm{\hat{W}}$ are 1x6 row vectors written as:
\begin{align*}
    \bm{\hat{T}}&=[\bm{\omega} ~\vert~ \bm{v}] &= [\bm{\omega} ~\vert~ \bm{r} \times \bm{\omega}]\\
    \bm{\hat{W}}&=[\bm{f} ~\vert~ \bm{m}] &= [\bm{f} ~\vert~ \bm{r} \times \bm{f}]
\end{align*}
where $\bm{\omega}$ is a unit angular velocity vector, $\bm{v}$ is a unit linear velocity vector, $\bm{f}$ is unit moment vector, $\bm{m}$ is unit force vector and $\bm{r}$ is the expression point of the screw. When, $\bm{\omega}$ and $\bm{v}$ are unitary vectors as described above, $\bm{\hat{T}}$ is called a unitary twist and similarly if $\bm{f}$ and $\bm{m}$ are unitary vectors, $\bm{\hat{W}}$ becomes a unitary wrench. The advantage of using unitary screws is that the mobility analysis can be performed by using just the geometric parameters of the involved surfaces.

This theory, initially developed for kinematic analysis of mechanisms \cite{Adams1999Application,gerbino2004investigate}, has been also widely applied to tolerance analysis \cite{Desrochers1998,Laperriere2000Tolerance} by the assumption of manufacturing and assembly defects are generally small displacements. The difference between these applications is that for mechanism analysis the inputs are large motions of one or more of the parts and the outputs are the rigid body displacements, velocities and forces; and in tolerance analysis the inputs are small variations due to the manufacturing or assembly process and the output are the small rigid body displacements and accumulated variations. For a mechanism model, the solution describes the motion regarding the time. For a static assembly, the tolerancing analysis gives the variation of the assembly regarding the nominal model \cite{elmaraghy_comprehensive_1998}.

The theory of screws is suitable to represent the $n$ degrees of invariance of a toleranced surface or degrees of freedom of a toleranced joint by concatenating in a matrix the set of $n$ twists describing each degree of invariance or freedom:
\begin{align*}
\mathcal{T} = \left[\begin{array}{c}    
                    \bm{ \hat{T_1} }\\ 
                    \bm{ \hat{T_2} }\\
                    ... \\
                    \bm{ \hat{T_n} }
\end{array}\right]
\end{align*}
$\mathcal{T}$ is usually called twist-matrix and by calculating its dual vector space, the corresponding wrench-matrix (also called reciprocal) can be obtained. The reciprocity of screws is one of the most important property of this theory since it allows to change easily from the twist-space to wrench-space and vice versa.

By computing the union of displacements it is possible to determine the mobility conditions of a couple of surfaces of a mechanical system \cite{Adams1999Application}. This can be performed by concatenating the respective twist-matrices $\mathcal{T}_1$ and $\mathcal{T}_2$ of the surfaces \cite{konkar_incremental_1995}:
\begin{align*}
Union(\mathcal{T}_1,\mathcal{T}_2) = \left[\begin{array}{c}    
                    \mathcal{T}_1 \\ 
                    \mathcal{T}_2
\end{array}\right]
\end{align*}
Returning to the example of figure \ref{fig:Case_study}, the twist representing the degree of invariance in rotation of $S_1$ and expressed in the global reference system $R_0$ is:
\[
\hat{T}_{11/0} = \left[\bm{y_{1/0}} ~~ \bm{0} \right]
\]
where $\bm{y_{1/0}}$ corresponds to $\bm{y_1}$ expressed in the global reference system $R_0$. Similarly, the twists expressed the global reference system $R_0$ representing the translations along $\bm{x_1}$ and $\bm{z_1}$ are:
\begin{align*}
    \hat{T}_{12/0} = \left[\bm{0} ~~ \bm{x_{1/0}} \right]\\
    \hat{T}_{13/0} = \left[\bm{0} ~~ \bm{z_{1/0}} \right]\\
\end{align*}

The twist-matrix $\mathcal{T}_{1/0}$ representing the degrees of invariance $S_1$ respect to $R_0$ is: 
\begin{align*}
\mathcal{T}_{1/0} = \left[\begin{array}{c}    
                    \hat{T}_{11/0} \\ 
                    \hat{T}_{12/0} \\ 
                    \hat{T}_{13/0}
\end{array}\right]
\end{align*}
The unbounded displacements in the positioning of $S_1$ respect to $S_2$ can be calculated as the union of the corresponding set of displacements:
\begin{align*}
\mathcal{T}_{S1/S2} = Union(\mathcal{T}_{1/0},\mathcal{T}_{2/0}) = \left[\begin{array}{c}    
                    \mathcal{T}_{1/0} \\ 
                    \mathcal{T}_{2/0}
\end{array}\right]
\end{align*}
Finally, by the reciprocal of the union of the displacements it is possible to calculate the bounded displacements:
\[
\mathcal{W}_{S1/S2} = reciprocal(\mathcal{T}_{S1/S2})
\]

Depending on the relative orientation of normal vectors $y_1$ and $y_2$ of $S_1$ and $S_2$, $\mathcal{W}_{S1/S2}$ can be composed by three wrenches (in the case of parallelism) or by one wrench (in the general case). In the first case, the three wrenches represent two rotations and one translation, and in the second case, the resulting wrench represents the rotation along the vector defined by $y_1 \times y_2$.

%

\section{Proposed approach}
\label{sec:Proposed}
The possible displacements of a toleranced surface inside its tolerance zone, or a toleranced joint inside its clearance, can only be analyzed and controlled along the bounded displacements of the nominal surface, i.e. the displacements which do not leave the nominal surface globally invariant. Moreover, when the displacements are analyzed respect to other surface, the number of controlled displacements can decrease according to the invariant geometric properties of the other surface. For instance, to describe the position of a spherical surface respect to a reference system, three parameters for describing three translations are required. If the position of the same spherical surface has to be described to respect a planar surface, just one parameter for controlling the displacement of both surfaces along the normal of the plane is needed.

Due to the afford mentioned, we propose in this work to apply the theory of screws to identify, before a Minkowski sum, the subspace of $\mathbb{R}^6$ in which the most of the displacements of the calculated polytope are bounded. By knowing this, it is possible to adapt the dimension of the operand polytopes by projecting them into the previously identified subspace and then to perform the Minkowski sum on it. Thus, the operations can be performed with polytopes in the smallest possible dimension instead of dealing with 6-polytopes. This is justified by the reduction on the complexity of the Minkowski sums when the dimension of the operands is decreased. 

\begin{figure}[htbp]
	\centering
		\includegraphics[width=9cm]{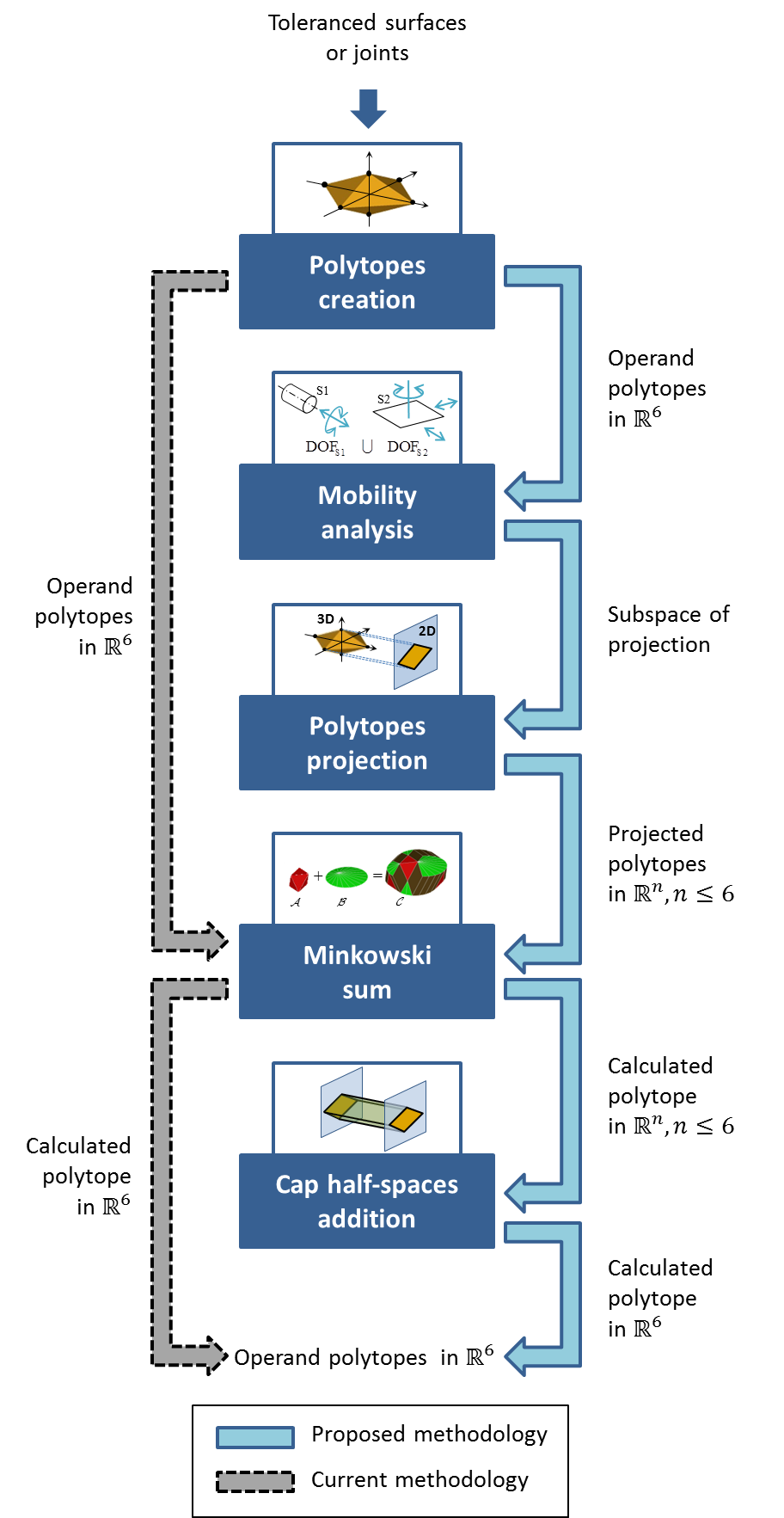}
	\caption{Proposed methodology vs current methodology}
	\label{fig:methodology}
\end{figure}

The proposed methodology consist in 5 steps as described following and depicted in figure \ref{fig:methodology}: 
\begin{enumerate}
\item Polytopes creation: for each geometric and contact tolerance a polytope has to be created. 
\item Mobility analysis: by the use of the theory of screws it is possible to determine the subspace of $\mathbb{R}^6$ containing the bounded displacements. In this subspace the operations has to be computed in order to deal with the simplest representation of the polytopes according to the tolerance analysis problem. 
\item Polytopes projection: when the subspace for operating the polytopes is identified, the projection of the operand polytopes is required. Some algorithms for projecting set of inequalities can be found in \cite{Cambridge2004Equality,Kebler1996Parallel}.
\item Minkowski sum: the projected polytopes can be summed by means of the software Politopix \cite{PolitoRef} in the identified subspace. The calculated polytope, representing the cumulative defects of the surfaces under control, will be the simplest representation of the tolerance analysis problem.
\item Cap half-spaces addition: in order to get a clear graphical representation of the calculated polytope in $\mathbb{R}^3$ the required half-spaces to limit the unbounded displacements have to be added. 
\end{enumerate}

%
\section{Case study}
The methodology proposed in previous section is illustrated in this section by solving the example presented in figure \ref{fig:Case_study}. Finally, a comparison with the solution by the 6-polytopes method given in section \ref{sec:six_polytopes} is presented.

The step (1) of the methodology is the same than in the case of the strategy with 6-polytopes. So, the next step consists on doing the mobility analysis to determine the displacements which allows the control of the position of $S_2$ respect to $S_1$. As explained in section \ref{sec:theory_of_screws}, the twist-matrix representing the unbounded displacements for positioning $S_2$ respect to $S_1$ is:
\[{{\mathcal T}_U} = \left[ {\begin{array}{*{20}{r}}
0&1&0&0&0&0\\
0&0&0&1&0&0\\
0&0&0&0&0&1\\
{sin(\theta )}&{cos(\theta )}&0&0&0&{ - d \cdot sin\left( \theta  \right)}\\
0&0&0&{cos(\theta )}&{sin(\theta )}&0\\
0&0&0&0&0&1
\end{array}} \right]\]
where $d$ is the distance along $y_0$ between the points $O_1$ and $O_2$.\\

Finally, the bounded displacements can be calculated as the reciprocal of $\mathcal{T}_U$ \cite{Adams1999Application}:
\[
\mathcal{W}_U = reciprocal(\mathcal{T}_U) = \left[ {0 ~~  0 ~~ 0 ~~ 0 ~~ 0 ~~ -1} \right]
\]

The previous result means that none force can be transmitted between the surfaces, but it is possible to transmit torque along $z_0$. From the tolerance analysis point of view, it means that for controlling the relative position of $S_2$ respect $S_1$ only to control the rotation along $z_0$, $r_{z_0}$, axis is required. Therefore, the subspace of $\mathbb{R}^6$ that contains just bounded displacement is a 1-dimensional space composed by $r_{z_0}$. Then, the Minkowski sum can be computed in $\mathbb{R}^1$ instead of in $\mathbb{R}^6$.

\begin{figure}[ht]
	\centering
		\includegraphics[width=8.5cm]{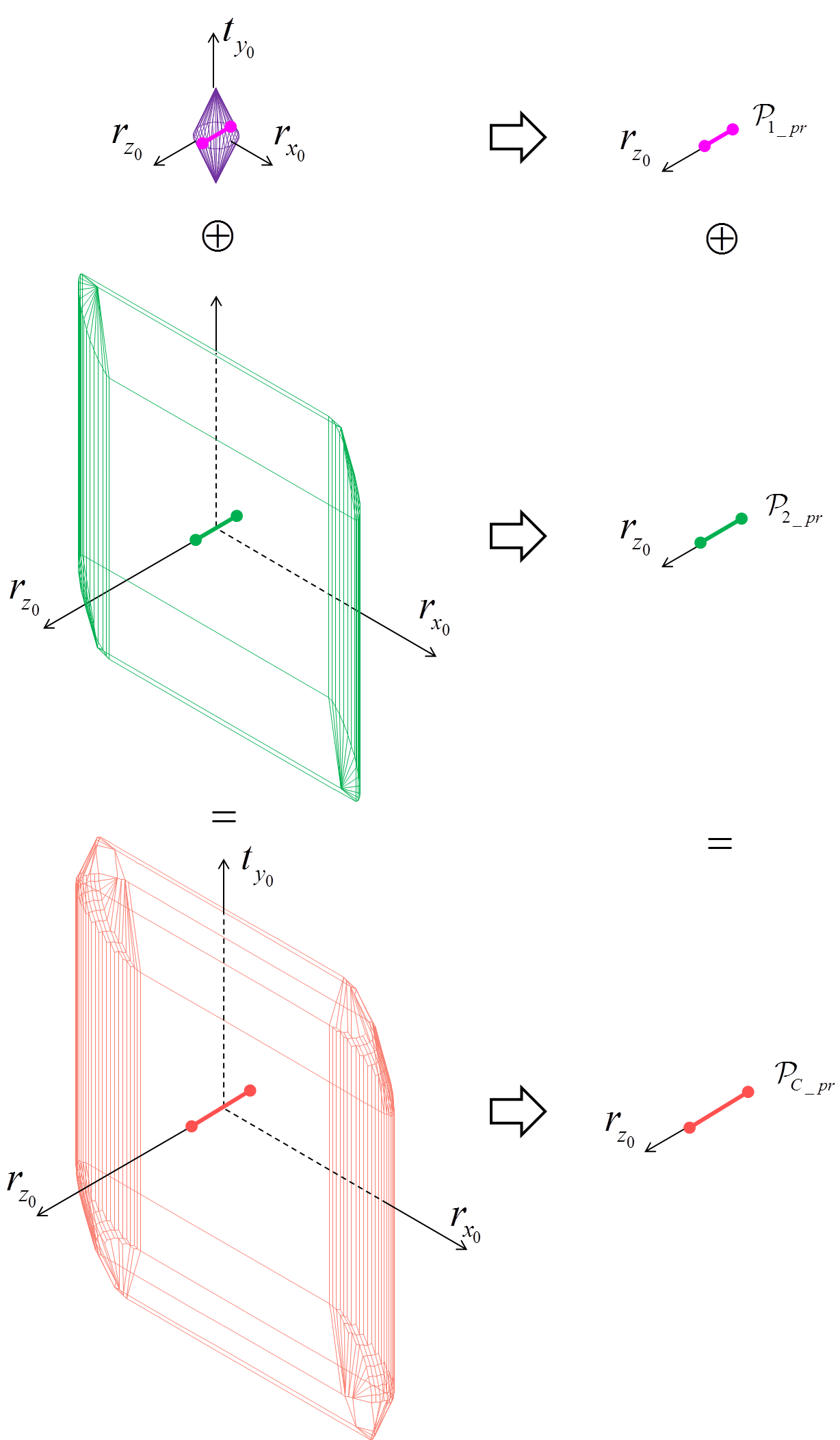}
	\caption{1D representation of $\mathcal{P}_{1\_pr}$, $\mathcal{P}_{2\_pr}$ and $\mathcal{P}_{C\_pr}$}
	\label{fig:sum1d}
\end{figure}

Next, according with step (3), the projections $\mathcal{P}_{1\_6D}\rightarrow \pi (\mathcal{P}_{1\_6D})=: \mathcal{P}_{1\_pr}$ and $\mathcal{P}_{2\_6D}\rightarrow \pi (\mathcal{P}_{2\_6D})=: \mathcal{P}_{2\_pr}$ were computed by means of ESP algorithm \cite{Cambridge2004Equality} into a 1-dimensional space that represents rotations along $z_0$ axis. $\mathcal{P}_{1\_pr}$ and $\mathcal{P}_{2\_pr}$ are therefore 1-polytopes composed by 2 facets and 2 vertices each one as it is summarized in table \ref{tab:1dresults}. Finally, the sum of $\mathcal{P}_{C\_pr} = \mathcal{P}_{1\_pr} \oplus \mathcal{P}_{2\_pr}$ in $\mathbb{R}^1$, also executed in politopix, took 0.001 s to be computed (in contrast with the 19 s of the initial method). The simulation was performed with an Intel Core i7-3740QM. Figure \ref{fig:sum1d} shows the calculated polytope and its operands in comparison with the 2D projection of its corresponding 6-polytopes.

\begin{table}[ht]
\centering
\caption{Tolerance analysis with projected polytopes  (F: facets, V: vertices)}
\label{tab:1dresults}
\begin{tabular}{llll}
\hline
             & $\mathcal{P}_{1\_pr}$  & $\mathcal{P}_{2\_pr}$  & $\mathcal{P}_{C\_pr}$ \\ \hline
F            & 2 & 2 & 2   \\
V            & 2 & 2 & 2   \\
Time {[}s{]} &   &   & 0.001   \\ \hline
\end{tabular}
\end{table}

\begin{figure}[ht!]
	\centering
		\includegraphics[width=6.4cm]{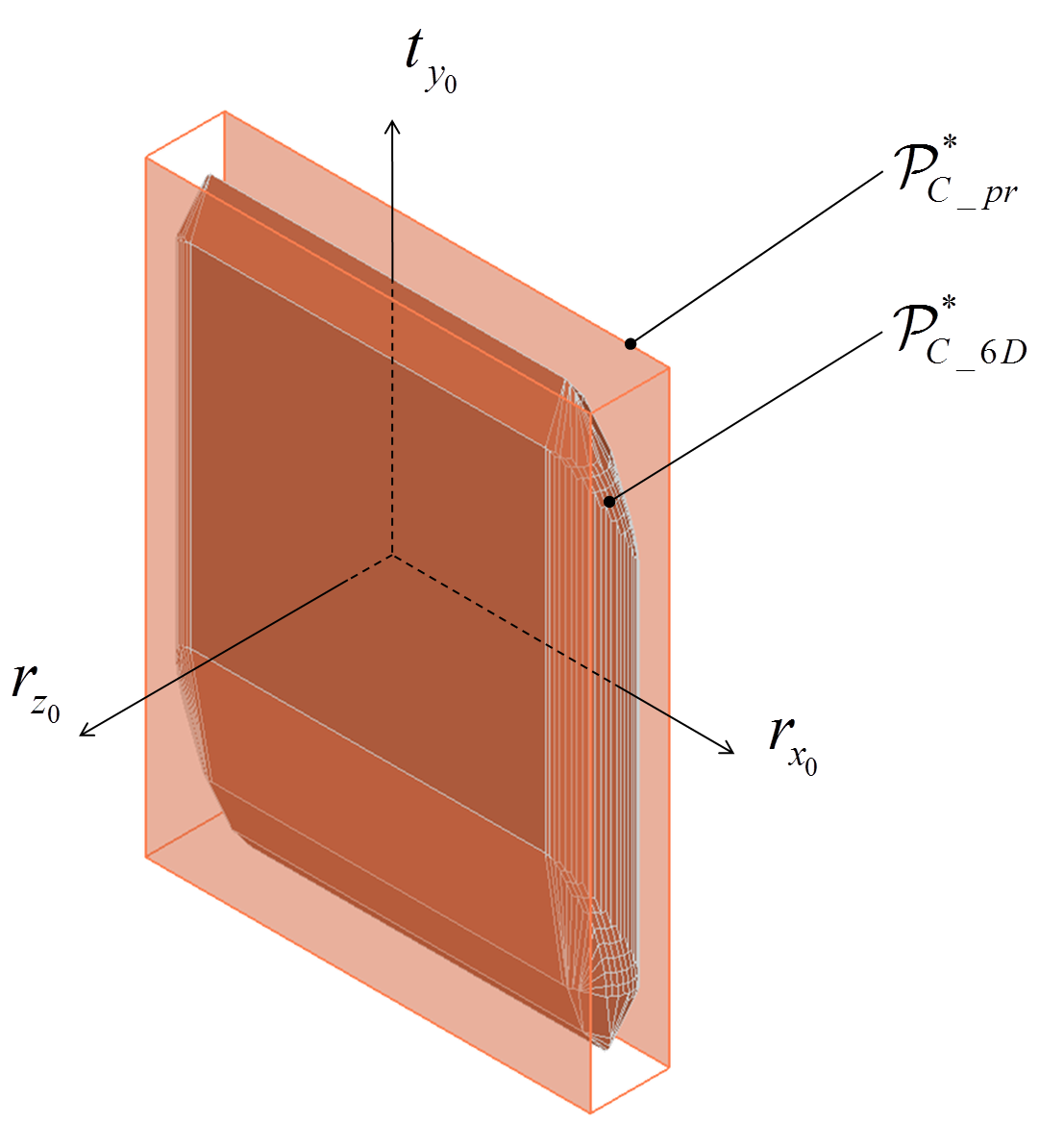}
	\label{fig:Pc_1D_vs_6D}
	\caption{Comparison of the 3D representation of the calculated polytopes $\mathcal{P}_{C\_6D}^*$ and $\mathcal{P}_{C\_pr}^*$}
\end{figure}

In order to check if the polytopes calculated by the two strategies $\mathcal{P}_{C\_pr}$ and $\mathcal{P}_{C\_6D}$ are equivalent together from the tolerance analysis point of view, the equality between $\mathcal{P}_{C\_pr}$ and the projection of $\mathcal{P}_{C\_6D}$ onto the subspace of the rotations along $z_0$ was checked. This was performed by evaluating if the vertices of $\mathcal{P}_{C\_pr}$ were inside of the half-spaces of the projection of $\mathcal{P}_{C\_6D}$ and vice versa. The equivalence can be also confirmed graphically in figure \ref{fig:sum1d}, where it can be noticed that the half-spaces of $\mathcal{P}_{C\_pr}$ agree with the half-spaces of $\mathcal{P}_{C\_6D}$ that are not caps (in figure \ref{fig:sum1d} cap facets are represented by dashed lines). The results can also be compared in figure \ref{fig:Pc_1D_vs_6D}, where it is presented a 3D representation of the result from both strategies. It can be noticed that $\mathcal{P}_{C\_6D}^*$ is composed by many cap facets coming from cap facets of the operands from which just the two bounding $r_{z_0}$ is required. It means that just 0.05\% of the facets of $\mathcal{P}_{C\_6D}$ represent useful information according to the tolerance analysis problem (see tables \ref{tab:6dresults} and \ref{tab:1dresults}). In the other hand, the 3D representation of the calculated polytope by the projection method $\mathcal{P}_{C\_pr}^*$ has the simplest topology to represent the associated tolerance analysis problem: a pair of non-cap half-spaces bounding $r_{_0}$, a pair of cap half-spaces bounding $r_{x_0}$ and a pair of cap half-spaces bounding $t_{y_0}$. This was achieved by identifying in advance the axes of bounded displacements and summing the projection of the operands in $\mathbb{R}^1$. This is the reason of the reduction in the computation time and the main contribution of the present work. 

\begin{figure}[t!]
	\centering
		\includegraphics[width=5.1cm]{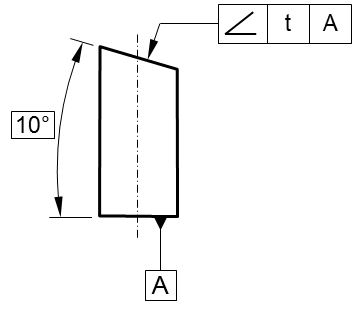}
		\caption{Special case of the case study when $TZ_1$ is zero}
	\label{fig:Part_case}
\end{figure}

In the particular case in which the tolerance for $S_1$ is zero, the analyzed case study corresponds to the ISO specification presented in figure \ref{fig:Part_case}. In fact, not Minkowski sum is required for this case and just the projection of the polytope of the toleranced surface on $\mathbb{R}^1$ is enough to obtain all the possible displacements between the two surfaces (i.e. the resulting polytope).
%

%
\section{Discussion and conclusions}

We proposed a way to calculate Minkowski sum of polytopes coming form tolerance analysis problems by adjusting the dimension of the operands according to the mobility conditions of the involved nominal surfaces. This with aim of avoiding computation of cap facets and deal as much as possible juts with half-spaces that represents real geometric constraints. Theory of screws is suitable for performing mobility analysis and determining in advance the subspace into which the operand polytopes have to be projected and summed to avoid calculating facets over the unbounded directions.

In comparison with the strategy based on 6-polytopes \cite{Homri2015103, Homri2015112}, the method proposed in this paper allows decreasing the computation time of Minkowski sums of polytopes by taking information from the tolerance analysis problem to simplify the operands and to perform the computation in the subspace of smallest possible dimension.

In some situations, as in the case of unilateral contacts, the absolute elimination of cap half-spaces is not possible, in other words, the set of contact constraints cannot be made compliant with a closed set in any subspace of $\mathbb{R}^6$. In such cases, it is required to use the virtual boundaries (cap half-spaces) introduced by Homri et al. \cite{Homri2015103, Homri2015112} and to trace them during the different computations in order to differentiate among all the facets of a calculated polytope between those that are generated by the cap half-spaces and the others generated by half-spaces that derive from geometric and contact constraints. The traceability of the vertices and facets of a calculated polytope from the vertices and facets of the operands represents an interesting direction for further research. By doing this, the constraints having more influence regarding the FC can be identified and then the maximization of the tolerances can be performed. Additionally, further research is required to generalize this method to solve complete tolerance analysis problems involving several parts.

\bibliographystyle{plain}
\bibliography{references}

\end{document}